\documentclass[pra,aps,superscriptaddress]{revtex4}
\usepackage{graphicx}
\usepackage{bm}
\usepackage{amsmath}
\usepackage{euscript}
\usepackage{graphics}
\usepackage{bm}
\usepackage{epsfig}
\usepackage{amssymb}

\begin{document}
\title{Coherent backscattering of light from saturated atoms}

\author{\firstname{Vyacheslav} \surname{Shatokhin}}
\affiliation{B.~I.~Stepanov Institute of Physics, National Academy
of Sciences, Skaryna Ave. 68, 220072  Minsk, Belarus}
\affiliation{Max-Planck-Institut f\"ur Physik komplexer Systeme,
N\"othnitzer Str. 38, 01187 Dresden, Germany}
\author{\firstname{Thomas} \surname{Wellens}}
\affiliation{Institut f\"ur Theoretische Physik, Universit\"at
Erlangen-N\"urnberg, Staudstr. 7, 91058 Erlangen, Germany}
\author{\firstname{Cord} \surname{M\"uller}}
\affiliation{Physikalisches Institut, Universit\"at Bayreuth, 95440
Bayreuth, Germany}
\author{\firstname{Andreas} \surname{Buchleitner}}
\affiliation{Max-Planck-Institut f\"ur Physik komplexer Systeme,
N\"othnitzer Str. 38, 01187 Dresden, Germany}

\begin{abstract}
We survey recent progress achieved in understanding the impact of
inelastic processes on coherent backscattering of light from cold
atoms that are saturated by a powerful laser field.
\end{abstract}

\maketitle
\section{Introduction}
\label{intro} 
\begin{figure}
\includegraphics[width=12cm]{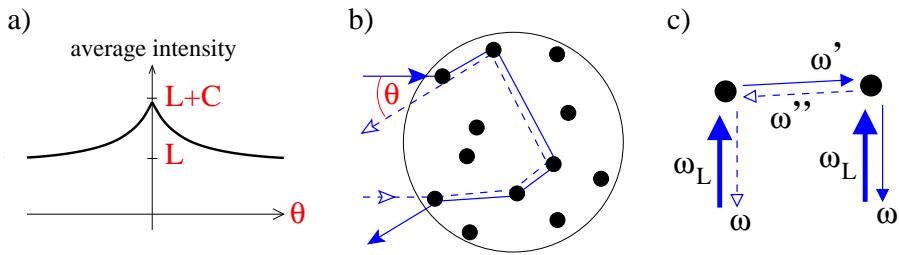}
\caption{(a,b) General mechanism of coherent backscattering:
  constructive interference between waves travelling along reversed
  scattering paths leads to an enhancement of the average
  intensity scattered from a disordered sample in backwards direction
  $\theta=0$. (c) In this paper, we examine the fundamental case of
two strongly laser driven atoms. Frequency fluctuations due to
  inelastic scattering break the symmetry between the reversed paths
  ($\omega'\neq\omega''$),
  leading to dephasing and reduction of the coherent backscattering peak.
} \label{fig:1}
\end{figure}

Coherent backscattering (CBS) of light is a dazzling
example of interference phenomena surviving a disorder average in
multiply scattering media \cite{meso94}. It occurs generally when
shining light onto an optically thick, disordered sample, provided
that the wave's phase coherence remains preserved over many
scattering events. In this so-called ``mesoscopic regime'',
interference effects lead to a complicated speckle pattern of the
wave intensity scattered into different directions. This pattern can
be seen as a fingerprint of the specific disorder realization. When
averaging over the disorder, however, most of the speckles are
washed out. The only interference peak non-sensitive to disorder is
the one in exact backscattering direction. It originates from
constructive interference between waves interacting with the same
scatterers, but in opposite order, see Fig.~\ref{fig:1}(b). In the
ideal case of perfectly constructive two-wave interference, the
peak-to-background ratio of backscattered intensities, known as the
enhancement factor $\alpha$ (see Fig.~\ref{fig:1}(a)), equals
exactly two, whereas $\alpha<2$ if decoherence or dephasing effects
are present.

CBS clearly shows that, in general,
wave propagation in disordered media cannot be fully described by a
simple diffusion equation. In the case of weak disorder, an
approximate diffusion model can nevertheless be maintained, provided
that the enhanced backscattering effect is accounted for by a
reduction of the diffusion constant (weak localization). For strong
disorder, i.e., if the mean free path becomes comparable to the wave length,
interference between reversed paths can even lead to a complete
absence of diffusion (strong localization) \cite{vollhardt}.
A promising candidate for
reaching the strong disorder regime experimentally, is the scattering
of light by cold atoms, which exhibit an extremely large resonant
cross section.

Since the first
experimental observation of CBS from cold atoms in 1999
\cite{labeyrie99}, numerous theoretical and experimental activities
have been devoted to elucidating relevant dephasing mechanisms, such
as Raman scattering on the degenerate atomic transitions
\cite{jonckheere00,mueller01,mueller02,kupriyanov03,labeyrie03,mueller05},
the influence of a magnetic field \cite{labeyrie02,sigwarth04},
thermal motion of atoms \cite{kupriyanov04,labeyrie06,wickles06},
and nonlinear response of saturated dipole transitions
\cite{chaneliere03,balik05,wellens04,shatokhin05,wellens05,wellens06,shatokhin06,gremaud06}.
Experiments have been performed on two atomic species, rubidium (Rb)
\cite{labeyrie99,labeyrie03,kupriyanov04,labeyrie06,balik05} and
strontium (Sr) \cite{chaneliere03,bidel02}.

In the present mini-review, we focus on the impact of inelastic
scattering from strongly driven, saturated atomic dipole transitions on CBS.
Understanding this dephasing mechanism, induced by quantum mechanical
frequency fluctuations of the scattered photons, is
important for the transition from weak to strong localization, which
is expected to occur at increased atomic densities, when atoms
exchange multiple photons, and even a single photon is able to
saturate the atomic transition \cite{chaneliere03}. Apart from its
fundamental interest, controlling this phase-breaking mechanism is
crucial for more technological applications such as random lasers
\cite{cao03}.

\section{The model}
\label{sec:2}
Considering a system of only two
atoms suffices to grasp the essential physical phenomena
\cite{wellens04,shatokhin05,shatokhin06,gremaud06}.
This relatively simple theoretical model describes an optically thin
atomic medium, where double scattering provides the dominant
contribution to the CBS signal \cite{jonckheere00,bidel02}. It is in
this double scattering regime that a CBS reduction due to the
saturation of atomic dipole transitions was first observed
\cite{chaneliere03}.

Our two-atom model system is described by the Hamiltonian
\begin{equation}
H=H_A+H_F+H_{AF}+H_{AL}. \label{eq:ham}
\end{equation}
Here, $H_A$ is the sum of the atomic Hamiltonians describing the
electronic dipole transitions of  identical, motionless atoms at
random positions ${\bf r}_1$ and ${\bf r}_2$, with the distance
$r=|{\bf r}_1-{\bf r}_2|$ much greater than the optical resonance
wavelength $\lambda$. The atomic internal structure corresponds to
the transition $J_g=0\leftrightarrow J_e=1$, precisely as in the Sr
experiment \cite{chaneliere03}. The atoms are coupled to the
quantized photon field bath $H_F$. The atom-field interaction
$H_{AF}$ leads to the radiative linewidth $2\gamma$ of the atomic
excited level. This interaction also permits the two atoms to
exchange resonant photons which implies a far-field dipole-dipole
coupling  of order $g\sim 1/k r$. Since we are considering the weak
localization regime $kr\gg 1$, the small coupling $g\ll 1$ implies
that atoms exchange only single photons with each other. However,
the atoms are exposed to an external driving laser field with
wavevector ${\bf k}_L$ and frequency $\omega_L$, the coupling being
described by the last term in Eq.~(\ref{eq:ham}).

Two important parameters describe the atom-laser interaction: the
detuning $\delta=\omega_L-\omega_0\ll\omega_0$ from the optical
resonance at transition frequency $\omega_0$, and the Rabi frequency
$\Omega$ describing the dipole coupling strength. The effective
laser intensity is conveniently described \cite{cohen_tannoudji} by
the saturation parameter $s=\Omega^2/2(\gamma^2+\delta^2)$. If $s\ll
1$, the atoms scatter photons elastically, while $s\simeq 1$
indicates the onset of inelastic scattering, where the frequency of
the scattered photons is different from the laser frequency.

\section{Results}
\label{sec:3}
\subsection{CBS intensity and enhancement factor}
\label{subsec:1} A quantity of primary interest is the CBS
enhancement factor $\alpha$ measuring the phase coherence between
counter-propagating waves:
\begin{equation}
\label{eq:enh} \alpha=\frac{L_2^{\rm tot}+C_2^{\rm tot}}{L_2^{\rm
tot}}.
\end{equation}
It is the total CBS intensity, measured at backscattering ${\bf
k}=-{\bf k}_L$, divided by the total background intensity $L_2^{\rm
tot}$ measured away from the backward direction, the index ``2''
indicating the double scattering contribution.
The radiated field amplitude is proportional to the electric dipole of
the emitting atom \cite{cohen_tannoudji}.
The scattered intensities, i.e., background $L_2^{\rm tot}$ and interference
$C_2^{\rm tot}$ contributions, can
therefore be expressed via certain dipole correlation functions and excited
state populations, respectively, of the two atomic CBS transition.

Figure~\ref{fig:1}(c) shows one example of the nonlinear inelastic
scattering processes, induced by a powerful laser field,
contributing to the CBS interference 
$C_2^{\rm tot}$. Note
that the intermediate photon frequencies of the counter-propagating
processes are, in general, different from each other, leading to the
violation of the reciprocity symmetry between the reversed paths,
 and hence to a decrease of $\alpha$.

Starting from the Hamiltonian (\ref{eq:ham}), we have derived
\cite{shatokhin05,shatokhin06} the two-atom master equation
governing the evolution of arbitrary two-atom observables. Thus, CBS
intensities and, hence, the enhancement factor can be found by
solving the master equations and performing a configuration average
of the results over the probability distributions of ${\bf r}_1$ and
${\bf r}_2$. In many cases, analytical solutions are found, while
general results are accessible from numerical solutions. In the
following, we will report results for the helicity preserving
polarization channel, referring the interested reader to
Ref.~\cite{shatokhin06} for more general cases.

\begin{figure}
\includegraphics[width=12cm]{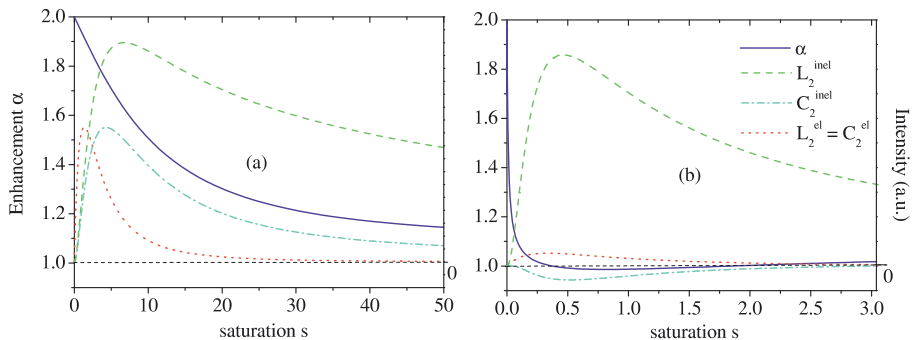}
\caption{Enhancement factor $\alpha$ (solid) vs. saturation $s$ for
(a) exact on-resonance driving ($\delta=0$), and (b) detuned driving
($\delta=20\gamma$). The dashed, dashed-dotted, and dotted lines
represent the inelastic background $L_2^{\rm inel}$, the inelastic
interference $C_2^{\rm inel}$, and the elastic $L_2^{\rm
el}=C_2^{\rm el}$ intensities, respectively. The fact that $\alpha$
approaches a limit $\alpha_{\infty}=23/21>1$ for $s\to\infty$
signals constructive photon (self-)interference in the deep
inelastic regime.}
\label{fig:2}       
\end{figure}

Both $C_2^{\rm tot}$ and $L_2^{\rm tot}$ can be decomposed into
elastic and inelastic components, $C_2^{\rm tot}=C_2^{\rm
el}+C_2^{\rm inel}$, $L_2^{\rm tot}=L_2^{\rm el}+L_2^{\rm inel}$.
Figure \ref{fig:2} shows the  elastic parts $L_2^{\rm el}=C_2^{\rm
el}$ (the equality holds thanks to reciprocity), the inelastic
background $L_2^{\rm inel}$ and interference term $C_2^{\rm inel}$
(right $y$ axis), as well as the enhancement factor $\alpha$ (left
$y$ axis), all as function of the saturation $s$ for two different
detunings: (a) on-resonance $\delta=0$ and (b) $\delta=20\gamma$.

The on-resonance enhancement factor in Fig.~\ref{fig:2}(a) decreases
linearly with small $s$, in qualitative agreement with the
experiment \cite{chaneliere03,note1}. In the highly saturated regime
$s\gg 1$, the elastic intensity is negligible, yet the limit value
$\alpha_\infty=23/21$ larger than unity demonstrates the residual
(self-)interference of inelastically scattered photons
\cite{shatokhin05}.

In the case of far-detuned driving shown in Fig.~\ref{fig:2}(b), an
enhancement $\alpha<1$ \emph{smaller} than unity around $s\simeq
1/2$ implies the presence of CBS anti-enhancement \cite{shatokhin06},
due to the destructive interference with $C_2^{\rm inel}<0$ of the
inelastic photons (cf.\ \cite{kupriyanov04} for a similar effect in
a different situation). The general condition for CBS
anti-enhancement in the saturation regime can be formulated as
$\Omega\simeq |\delta|\gg \gamma$, which corresponds to $s\simeq
1/2$. Also in the far-detuned case, $\alpha$ tends to a limit value
as $s\to\infty$, which is larger than unity but smaller than
$\alpha_\infty$.

\subsection{CBS spectrum} \label{subsec:2}
The results shown in Sec.~\ref{subsec:1} clearly demonstrate that
inelastic photons do contribute to the CBS interference. It is
therefore interesting to resolve the spectral characteristics of the
background and interference contributions. In other words, instead
of measuring the total intensities defining $L_2^{\rm tot}$ and
$C_2^{\rm tot}$, we wish to detect the backscattered light by a
narrowband detector tuned to a frequency $\omega$. Such a set-up,
completely in the spirit of Fig.~\ref{fig:1}(c), enables one to probe
the elementary scattering processes contributing to the CBS signal at
the frequency $\omega$. By varying $\omega$, it is possible to obtain a
more detailed characterization of the inelastic processes having impact
on the CBS signal in the saturation regime.

\begin{figure}
\includegraphics[width=12cm]{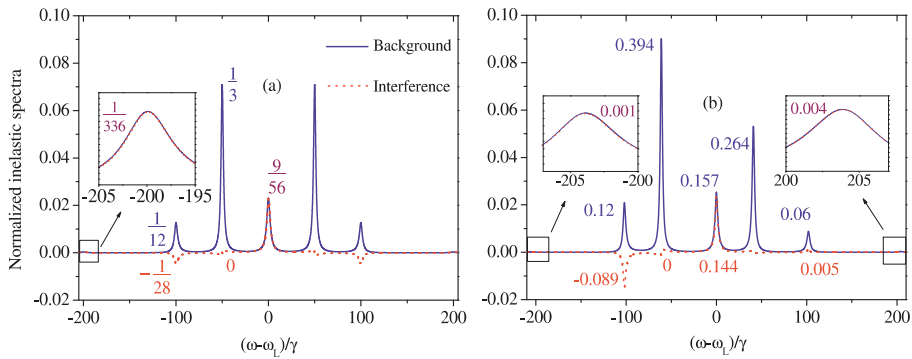}
\caption{Normalized inelastic spectra of the background (solid) and
interference (dotted) terms, in the limit of well separated spectral
lines, at $\Omega=100\gamma$. (a) $\delta=0$; (b) $\delta=20\gamma$.
The numbers near the resonances indicate their areas, such that the
overall areas of the background and interference terms give unity
and $C_2^{\rm inel}/L_2^{\rm inel}$, respectively. This corresponds
to (a) $\alpha=\alpha_\infty\simeq 1.096$; (b) $\alpha=1.065$.}
\label{fig:3}
\end{figure}

In Fig.~\ref{fig:3}, we
present the CBS spectrum in the limit of well separated spectral
lines of the single-atom resonance fluorescence spectrum
($\Omega\gg\gamma$). These results were calculated (analytically
for exact on-resonance, and numerically for detuned driving) with
the same master equation approach in combination with the quantum
regression theorem \cite{shatokhin07,shatokhin07b}.

In the limit $\Omega\gg\gamma$, the elastic intensity vanishes, and
the CBS spectrum is purely inelastic (see Fig.~\ref{fig:3}). Both
the background and interference spectra consist of seven resonances.
The resonances of the background spectra are Lorentzians with
positive weights, defining areas of the respective peaks in
Fig.~\ref{fig:3}. As for the interference contribution, its spectrum
represents a combination of the Lorentzian peaks with positive and
negative weights as well as dispersive resonances. The total areas
of the interference resonances, $2/21$ for $\delta=0$ and $0.065$
for $\delta=20\gamma$, allow to deduce the values of the CBS
enhancement factor, $\alpha=23/21$ and $\alpha=1.065$, respectively.

\begin{figure}
\includegraphics[width=12cm]{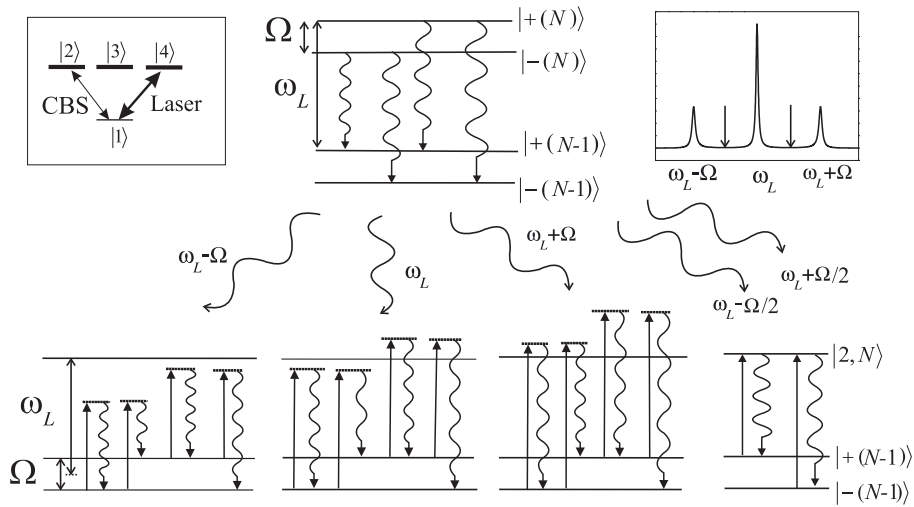}
\caption{Internal structure of the laser driven and CBS transitions
of the atoms. Spontaneous emission processes between dressed states
(top, center) give rise to the Mollow triplet (top, right) with
peaks centered at $\omega_L-\Omega$, $\omega_L$, and
$\omega_L+\Omega$. Apart from these, one needs to take into account
photons with frequencies $\omega_L\pm\Omega/2$, indicated by the two
downward arrows in the plot of the Mollow triplet. The photons
emitted by atom 1 (wiggly arrows with indicated frequencies)
propagate to atom 2 (bottom), where they are (re-)scattered from
either of the dressed states
 $|+(N-1)\rangle$, $|-(N-1)\rangle$, finally giving rise to
seven different peaks in the two-atom spectrum
(Fig.~\ref{fig:3}).} \label{fig:4}
\end{figure}

The number and positions of the CBS spectral peaks can be understood
within a dressed-state analysis of the atomic laser-driven and CBS
transitions (see Fig.~\ref{fig:4}) \cite{note2}. We assume a right
circular polarization of the laser field driving the
$|1\rangle\leftrightarrow |4\rangle$ transition of both atoms. Then,
observing CBS in the helicity preserving channel corresponds to
detecting photons emitted from the $|1\rangle\leftrightarrow
|2\rangle$ (Fig.~\ref{fig:4}, top left).

The interaction of the atoms with the powerful laser field leads to the
formation of dressed states \cite{cohen_tannoudji}
\begin{equation}
|\pm(N)\rangle=2^{-1/2}(|1,N+1\rangle\pm|4,N\rangle),
\end{equation}
where $N$ is the number of photons in the laser mode
(Fig.~\ref{fig:4}, top center). Spontaneous transitions from the
dressed-state manifold $\{|\pm(N)\rangle\}$ to
$\{|\pm(N-1)\rangle\}$ give a resonance fluorescence spectrum with
three peaks centered at $\omega_L-\Omega$, $\omega_L$, and
$\omega_L+\Omega$, which is known as the Mollow triplet
\cite{mollow69} (Fig.~\ref{fig:4}, top right). Photons emitted by
one atom are re-scattered on the CBS transition of the other atom.
The level $|2\rangle$ of the latter transition is not affected by
the laser laser field. However, the
$|1\rangle\leftrightarrow|2\rangle$ transition is modified by the
laser field, because it shares the common level $|1\rangle$ with the
laser-driven transition. Therefore, the internal structure of the
CBS transition is such as shown in Fig.~\ref{fig:4}(bottom): it has
one excited state and two ground state sublevels separated by
$\Omega$. Correspondingly, the new resonance frequencies of the CBS
transition are $\omega_L\pm\Omega/2$.

When the Mollow triplet emitted by one atom is incident on another
atom, it is scattered on the internal structure of the latter.
The relevant scattering processes that can take place are depicted on
the bottom of Fig.~\ref{fig:4}. Each photon can be scattered either
elastically or undergo Raman-Stokes or -anti-Stokes (multiphoton)
transitions (which lead to a frequency change by $-\Omega$ or
$\Omega$, respectively), which conserve energy and angular momentum.
It follows that the CBS spectrum must have resonances at
$\omega=\omega_L\pm 2\Omega$, $\omega=\omega_L\pm\Omega$, and
$\omega=\omega_L$. The diagrams describing the emission of CBS
photons at these frequencies are the three left-most diagrams on the
bottom of Fig.~\ref{fig:4}. The right-most diagram in
Fig.\ref{fig:4}(bottom) depicts resonant scattering of photons
with frequencies $\omega_L\pm\Omega/2$, which leads to an additional
doublet -- the Autler-Townes doublet \cite{townes55} -- in the CBS
spectrum.

The above analysis can straightforwardly be extended to the general
case $\delta\neq 0$. Again, the background and interference spectra
both consist of seven resonances (see Fig.~\ref{fig:3}(b)) which
represent: (i) a (re-)scattered Mollow triplet \cite{mollow69} at
$\omega=\omega_L,\, \omega_L\pm\Omega^{\prime}$, where
$\Omega^{\prime}=(\Omega^2+\delta^2)^{1/2}$,  (ii) an Autler-Townes
doublet \cite{townes55} at $\omega=\omega_L\pm
(\Omega^\prime\mp\delta)/2$, and (iii) a doublet at
$\omega=\omega_L\pm2\Omega^\prime$ originating from Raman-Stokes and
anti-Stokes scattering of the Mollow triplet sidebands .

Whether the interference of amplitudes at a certain frequency is
constructive or destructive is determined by their relative
frequency-dependent phase
shifts \cite{shatokhin07b}. The interference is purely constructive around
$\omega=\omega_L$ and $\omega=\pm 2\Omega^\prime$, whereas it
changes its character around the dispersive resonance peaks of the
Autler-Townes doublet. Finally, the interference can be destructive for
one or both of the Mollow sidebands (see Fig.~\ref{fig:3}). At
$|\delta|\simeq \Omega$, these destructive interferences can
outweigh the constructive ones in the sum over all spectral
contributions, resulting in CBS anti-enhancement for the total
intensity.

\section{Conclusion and outlook}
\label{sec:4} We have studied the impact of inelastic processes on
CBS  within a simple quantum optical model involving two atoms
exposed to a strong laser field. Within our master-equation approach, we
can calculate the loss of CBS interference for all values of saturation
and detuning, including an analysis of the spectral components.
 A dressed-state picture permits to understand the
position and character of resonance peaks that constitute the
background and interference signals.

Including higher scattering orders as well as addressing propagation
effects in bulk atomic clouds is hard within this framework, due to
the exponential growth of the Hilbert space with the number of
scatterers. A promising direction of future research is to unify the
presently discussed master-equation description of atom-photon
interaction with the diagrammatic scattering approach  that has been
developed for nonlinear classical scatterers
\cite{wellens05,wellens06}. Further challenging problems include a
quantitative explanation of the CBS experiment in the saturation
regime with internally degenerate Rb atoms \cite{balik05}, the
assessment of quantum statistical properties of the backscattered
field, and, ultimately, the exploration of the strong localization regime $kr\simeq
1$.


\begin{thebibliography}{99}
\bibitem{note1}
This loss of interference visibility can be traced back to the
availability of which-path information encoded in inelastically
scattered photons \cite{wellens04}; see also the similar situation
for thermal motion \cite{wickles06} and internal degeneracy
\cite{miniatura07}.
\end{thebibliography}
\end{document}